\documentclass[USenglish]{article}	
\usepackage[utf8]{inputenc}				%(only for the pdftex engine)
\usepackage[big,online]{dgruyter}	%values: small,big | online,print,work
\usepackage{lmodern} 
\usepackage{bm} 
\usepackage{microtype}
\usepackage{upgreek}
\usepackage[numbers,square,sort&compress]{natbib}

\theoremstyle{dgthm}

\theoremstyle{dgdef}

\begin{document}

%%%--------------------------------------------%%%
	\articletype{Research Article}
	\received{Month	DD, YYYY}
	\revised{Month	DD, YYYY}
  \accepted{Month	DD, YYYY}
  \journalname{De~Gruyter~Journal}
  \journalyear{YYYY}
  \journalvolume{XX}
  \journalissue{X}
  \startpage{1}
  \aop
  \DOI{10.1515/sample-YYYY-XXXX}
%%%--------------------------------------------%%%

\title{Instrumentation and Stabilization of Electric Arcs for Plasma Smelting Reduction}
\runningtitle{Plasme Arc Stabilization}
%\subtitle{Insert subtitle if needed}

\author*[1]{Thomas Voglhuber-Brunnmaier}
\author[2]{Lukas Ecker}
\author[3]{Benjamin Lehner}
\author[4]{Jakob Stanek}
\author[5]{Erwin K. Reichel} 
\runningauthor{T.~Voglhuber-Brunnmaier et al.}
\affil[1]{\protect\raggedright 
Institute for Microelectronics and Microsensors, Johannes Kepler University, Linz, Austria, e-mail: thomas.voglhuber-brunnmaier@jku.at}
\affil[2]{\protect\raggedright 
Institute of Control Systems, Johannes Kepler University, Linz, Austria, e-mail: lukas.ecker@jku.at}
\affil[3]{\protect\raggedright 
Institute for Microelectronics and Microsensors, Johannes Kepler University, Linz, Austria}
\affil[4]{\protect\raggedright 
Institute for Microelectronics and Microsensors, Johannes Kepler University, Linz, Austria}
\affil[5]{\protect\raggedright 
Metallurgical Competence Center, K1-met, Stahlstraße 14, 4020 Linz }
	
%\communicated{...}
%\dedication{...}
	
\abstract{The Hydrogen Plasma Smelting Reduction (HPSR) is a promising technology for low-CO$_2$ steel production. However, unstable arc conditions limit energy efficiency, hydrogen utilization, and electrode wear. In this paper, we discuss the influence of the electrical supply characteristics, which can be a source of additional instability. For active stabilization, a fast current controller using a buck converter with an adaptive state controller is used. For analyzing arc stability, a stereoscopic measurement system for reconstructing the arc is presented. The results show a clear correlation between the arc lengths and the measured arc voltage at constant current. Therefore, arc stability can be determined by the measured voltage. These findings build the basis for dynamic arc control by fast electrical interventions.}

\keywords{buck converter, stereoscopy, adaptive state controller, steel making}

\maketitle

\section{Introduction} 
The hydrogen-plasma smelting reduction (HPSR) process is one of the most promising approaches towards decarbonized steel production~\cite{Hiebler,Pauna,Adami}. Fine ores are reduced in a single step and melted to separate gangue and molten crude steel. The reduction of the iron ore is not achieved by adding carbon, but by hydrogen ionized in an electric arc. In this way, both the introduction of melting energy and the provision of the reducing agent can be provided entirely by renewable energy sources. In the target process, iron ore and a mixture of argon and hydrogen are supplied through hollow graphite electrodes. The argon promotes plasma formation. In the literature ~\cite{Zarl1,Leitner,Zarl2}, it is stated that unstable plasma conditions, i.e., fluctuations in arc lengths up to temporary arc extinction, can lead to unusually high wear of the graphite electrodes and to reduced electrical efficiencies and poorer hydrogen utilization. Said plasma instabilities result, on the one hand, from the complex coupling of thermodynamic and electrical forces, but also from the interaction of the negative differential conductivity of the arc and the characteristics of the electrical supply circuit~\cite{Reichel}. This last aspect, which has so far been largely ignored in the literature, is given special attention in this work. The aim is to investigate hidden dynamics of the arc and to stabilize it through rapid electrical intervention. 
\begin{figure}[htbp]
	\centering
	\includegraphics[width=0.35\textwidth]{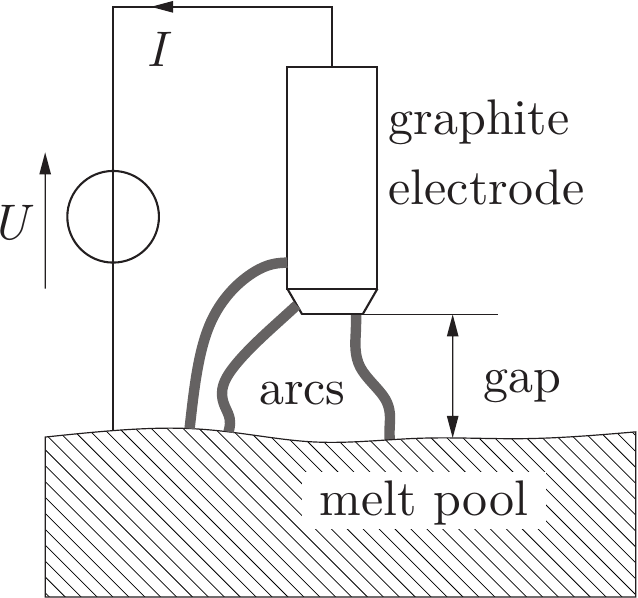}\hspace{1cm}
    \includegraphics[width=0.5\textwidth]{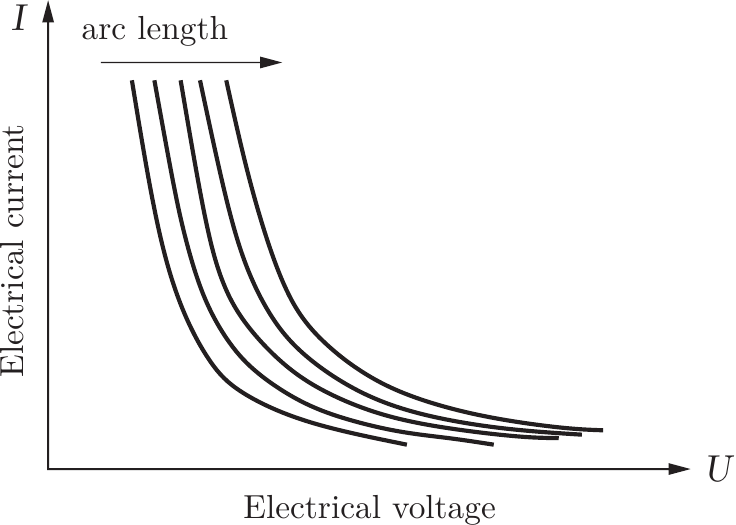}
	\caption{Left) Three examples of arcs. The arc tends to follow the electric field, but thermodynamic and electromagnetic effects cause the arc to dynamically deviate from the shortest path.) Right) Typical UI characteristics of arcs of different lengths. Increasing the gap qualitatively shifts the curves to the right.} \label{fig:fig1}
\end{figure}
Figure~\ref{fig:fig1} (left) shows an illustration of the electrode over the smelting pool. The charged particles of the plasma follow the electric field, preferably along the shortest path contacting the electrode and the melt pool in normal directions. However, this situation is rarely the case, as thermodynamic and electromagnetic effects cause nonstationary arc conditions. 
Figure~\ref{fig:fig1} (right) shows three arcs of different lengths. The negative differential resistance is the consequence of increasing ionization with increased power input. Therefore, with increasing current, the conductivity increases, which reduces the voltage drop. Negative differential resistance potentially leads to instabilities of the supply circuit, provided that supply line inductance and capacitance are taken into account~\cite{Rieder}. Extension of the arc with a constant current causes an increase in voltage. The erraticity of the arc (i.e., the stochastic arc length variation)  is influenced by many factors, including gas composition, flow rates, and temperatures, but also by the electrical power input. The hypothesis is that dynamic arc length changes should be correlated with the arc voltage at constant current. As a consequence, the plasma voltage variance could be used as a control error function for a supervisory controller, which is, however, not within the scope of this work.  
\begin{figure}[htbp]
	\centering
	\includegraphics[width=0.49\textwidth]
	{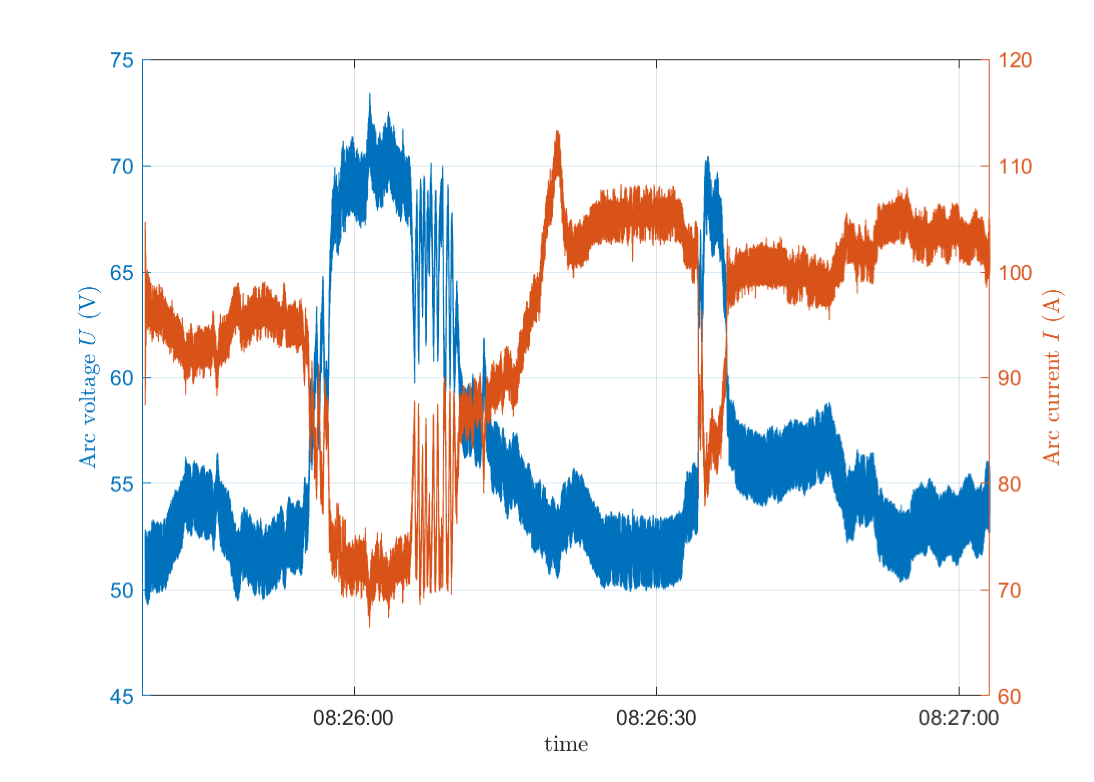}
    \includegraphics[width=0.49\textwidth]
	{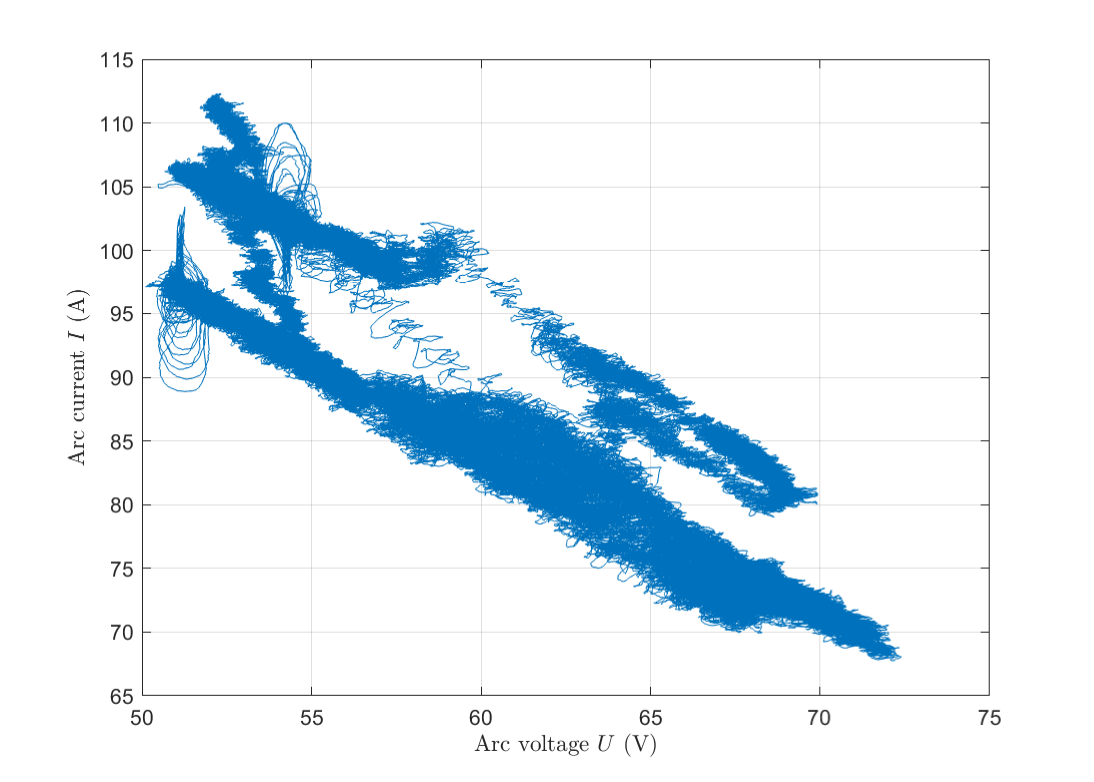}
	\caption{Left) Raw voltage and current data from an arc test over time. Right) current versus voltage of the same data.} \label{fig:fig2}
\end{figure}
Figure~\ref{fig:fig2} shows time series of the measured voltage and current in experiments on an existing laboratory system designed to investigate metallurgical properties of the smelting reduction process. The electrical signals were not of interest in the original investigations. The supply circuit consisted of a thyristor-based phase-angle controller connected to a three-phase transformer and a subsequent 6-pulse rectifier. The operating point was set by adjusting the phase angle of the thyristor controller. The exemplary voltage and current curves in Fig.~\ref{fig:fig2} (left) are relatively erratic. The corresponding UI diagram in Fig.~\ref{fig:fig2} (right) indicates sections with stable arc lengths. To identify these sections, block-by-block estimation of the differential resistance $r_D$ and the (fictitious) short-circuit current $I_0$ according to
\begin{equation}
    I=\frac{U}{r_D}+I_0,\label{eqn:eqn1}
\end{equation}
was performed, and the residuals determined by 
\begin{equation}
\text{res}=\sum\limits_{i} \left|I-\frac{U}{\hat{r}_D}-\hat{I}_0\right| \label{eqn:eqn2}
\end{equation}
were analyzed. 
\begin{figure}[htbp]
	\centering
	\includegraphics[width=0.45\textwidth]
	{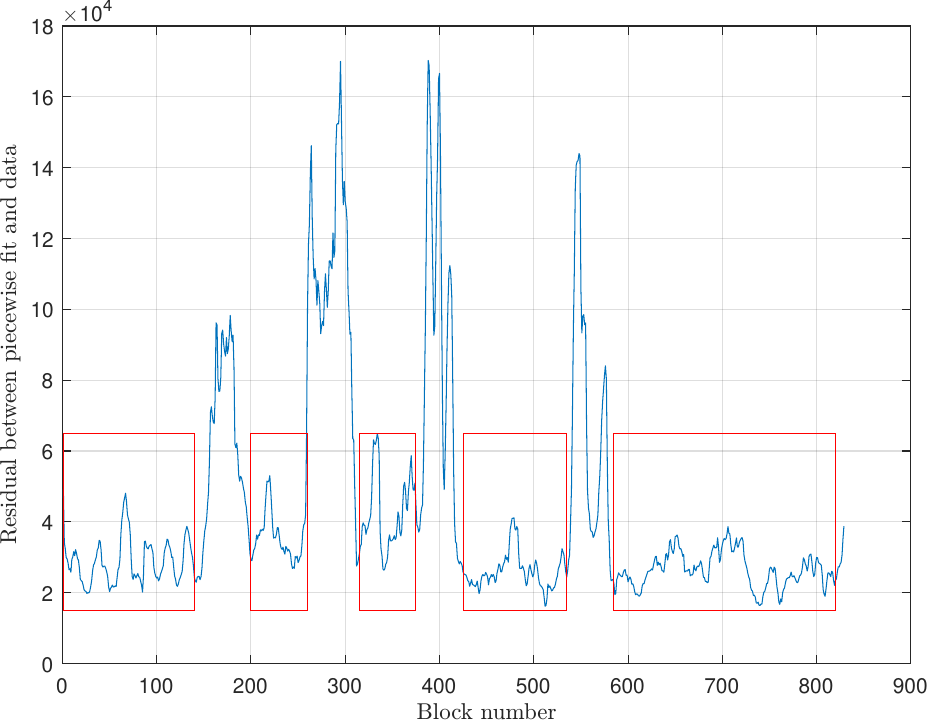}
    \includegraphics[width=0.51\textwidth]
	{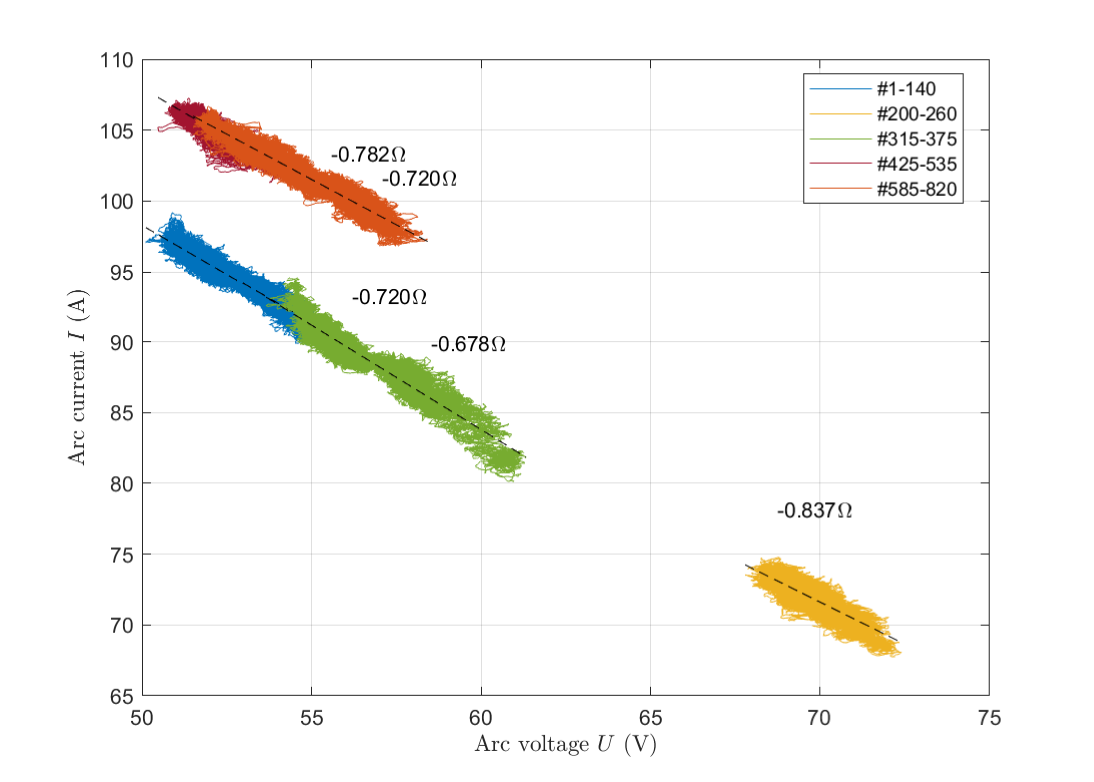}
	\caption{Left) Residuals of the block-wise estimates of $r_D$ and $I_0$ (window length is $10^5$ samples, i.e., 1s; Block displacement is 1/10s). Sections outlined in red are evaluated. Right) The selected measurement data reproduce the arc characteristic curves. The local differential resistances were calculated individually for the 5 blocks shown on the left.} \label{fig:fig3}
\end{figure}
Domains with low values (red boxes in Fig.~\ref{fig:fig3}(left)) indicate stable conditions, whereas high values are associated with arc transitions, e.g., a change in the point of impact on the melt pool or a change in the set-point. The diagram of the stable segments in Fig.~\ref{fig:fig3} (right) suggests that during most of the operation, the arc jumped between two metastable lengths and that the variations in time signals were mainly caused by the unstable supply circuit. Therefore, an unstable supply will be a source of instability, which will also influence the metallurgy. 
\section{Theory}
\begin{figure}[htbp]
	\centering
	\includegraphics[width=0.60\textwidth]{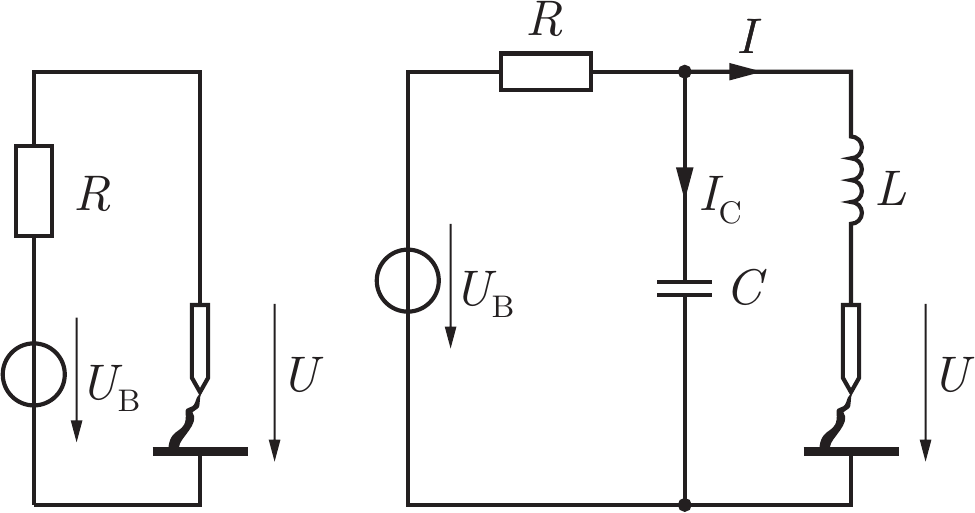}
	\caption{Left) Static equivalent circuit of source and arc. Right) Equivalent circuit with spurious $L$ and $C$ components influencing arc stability.} \label{fig:fig4}
\end{figure}

An electric arc is a complex electro-hydrodynamic phenomenon in which electrical, thermal, fluid-dynamic, and electromagnetic processes are coupled. The well-known erratic behavior of the arcs is mainly due to turbulence caused by the buoyancy of heated gas reaching tens of thousands of K, and the Lorentz-force interaction of charged particles. Many other effects come into play, but to illustrate the interaction of the electrical power supply and the arc, we first consider a stationary arc fed by a voltage source $U_\text{B}$ with some internal resistance $R$, shown in Fig.~\ref{fig:fig4} (left).
\begin{figure}[ht]
	\centering
	\includegraphics[width=0.45\textwidth]	{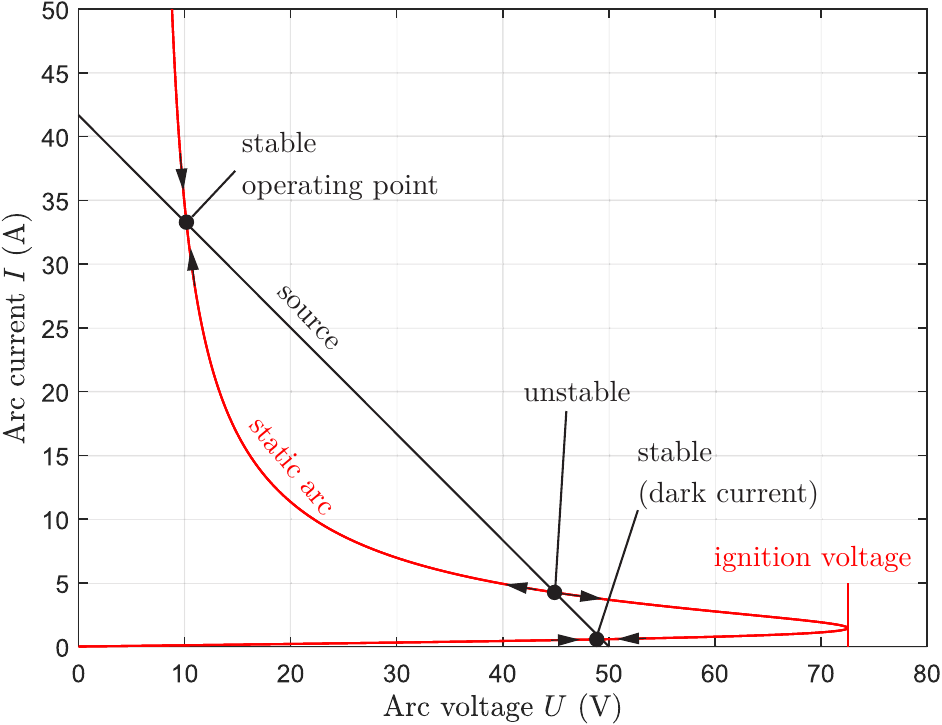}
    \includegraphics[width=0.45\textwidth]	{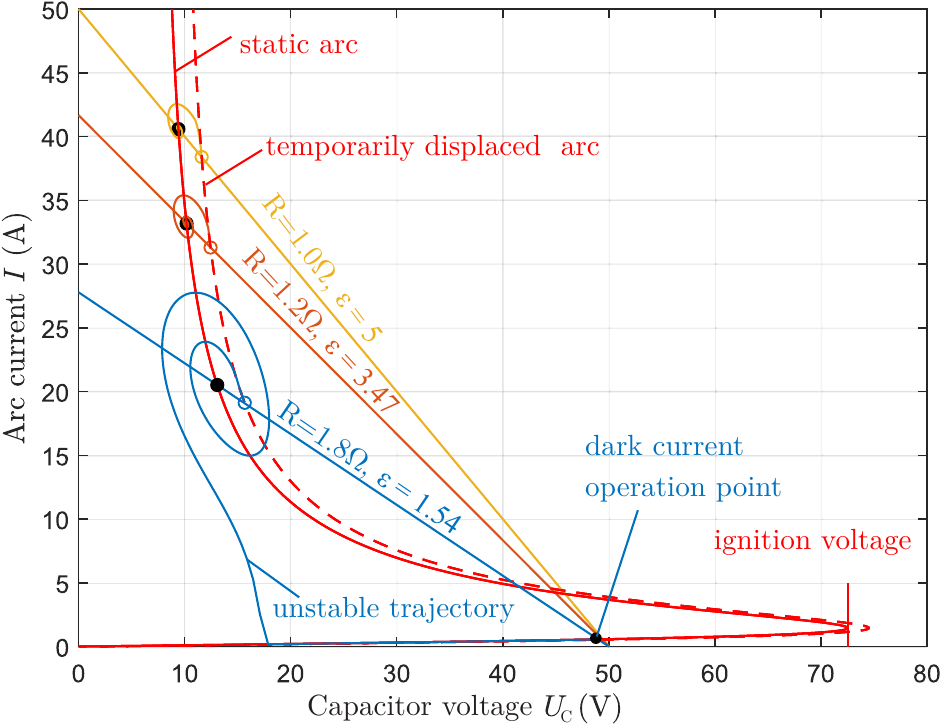}
	\caption{Left) Three potential operating points for a static arc connected to a source with internal resistance. Right) In the presence of reactive spurious elements $R$ and $L$ temporary arc excursions can cause oscillations and arc extinction.} \label{fig:fig5}
\end{figure}
Fig.~\ref{fig:fig5} (left) shows the three potential operating points for an illustrative electric arc. The used phenomenological arc model in Eq.~\ref{eqn:sjdfklgs1} is found in~\cite{Andrea} and is also used in~\cite{Lukas}
\begin{equation}
    U=\frac{\alpha R_0 I}{\arctan(\beta I)R_0 I+\alpha}.\label{eqn:sjdfklgs1}
\end{equation}
Here, $\alpha=10.5$\,V, $\beta=0.05$\,A$^{-1}$ and $R_0=100\,\Omega$ are parameters chosen for a clearer illustration, but do not correspond to the actual values of the experiments shown later. When the voltage is switched on, only a negligible dark current flows. Decreasing the gap such that the ignition voltage falls below the supply voltage, the arc ignites and the stable operating point at high current and low voltage results. The condition for stability of an operating point (o.p.) can be shown to be given by
\begin{equation}
    -\left.\frac{d U}{d I }\right|_\text{o.p.}<R,\label{eqn:sdfsdjfhgsd}
\end{equation}
which means that the negative resistance of the arc at the operating point has to be overcompensated by the source resistance for stability. Consequently, a current source is more suitable for generating electric arcs than a voltage source. If spurious elements such as output capacitor and wire inductance are considered (see Fig.~\ref{fig:fig4} (right)), stability is more intricate. A corresponding equivalent circuit as given in~\cite{Rieder} is shown in Fig.~\ref{fig:fig5} (right). In the following, it is assumed that some force elongates the arc, shifting the static curve temporarily to the right (see Fig.~\ref{fig:fig1} (right)). Whether the system state returns to the equilibrium point is analyzed using the nonlinear ordinary differential equations obtained from Kirchhoff’s laws:
\begin{align}
    U_\text{B}&= (I+I_C)R+U_C,\\
    I_C&=\frac{d U_C}{dt}C,\nonumber\\
    U_C&=\frac{d I }{dt}L+U(I).\nonumber
\end{align}
By introducing the normalized quantities
\begin{align}
\tau=\frac{t}{R C},\, u_c= \frac{U_C}{U_B},\, u= \frac{U(I)}{U_B},\, i = \frac{I R}{U_B},
\end{align}
the characteristic ordinary  differential equation (ODE) system is obtained
\begin{align}
\varepsilon\frac{d i}{d\tau}&=u_c-u(i),\label{eqn:sdflsödkfjg}\\
\frac{du_c}{d\tau}& = 1-i-u_c,\nonumber
\end{align}
with dimensionless characteristic parameter $\varepsilon$ defined by
\begin{align}
\varepsilon=\frac{L}{R^2C}.
\end{align}
The ODE system is in the form of a fast-slow system, where $\varepsilon$ determines which variable is the ``faster'' in the phase diagram in Fig.~\ref{fig:fig5} (right). For $\varepsilon>1$ the voltage $u_c$ reacts faster than the current $i$ in the phase diagram showing $i$ over $u_c$. The dimensionless parameter $\varepsilon$ is also related to the quality factor of a parallel RLC circuit by $\varepsilon=Q^{-2}$. The eigenvalues of the Jacobi matrix to Eq.~\ref{eqn:sdflsödkfjg} determine the local stability of an equilibrium point. I.e., the zeros $z_{1,2}$ of the characteristic polynomial~\cite{Strogatz}
\begin{align}
    p=\left.\text{det}\left(\begin{bmatrix}
        \frac{\rho_D}{\varepsilon}-z & \frac{1}{\varepsilon}  \\
        -1 &-1-z
    \end{bmatrix}\right)\right|_\text{o.p.}\rightarrow z_{1,2}= -\frac{1}{2}+\frac{\rho_D}{2\varepsilon} \pm \sqrt{\frac{1}{4}+\frac{\rho_D}{2\varepsilon}\left(1+\frac{\rho_D}{2\varepsilon}\right)-\frac{1}{\varepsilon} }\label{eqn:sdlskdjf2}
.\end{align}
Here, $\rho_D=-du/di$ denotes the negative differential normalized arc resistance. For local asymptotical stability (i.e. real parts of $z_{1,2}$ are both negative) a condition additional to Eq.~\ref{eqn:sdfsdjfhgsd} for $\varepsilon$ and $\rho_D$ is
\begin{align}
   \varepsilon>\rho_D \rightarrow
   \frac{L}{R C} > -\left.\frac{d U}{dI}\right|_\text{o.p.}\label{eqn:skljshdf}.
\end{align}

Figure~\ref{fig:fig5} (right) shows the exemplary trajectories for small excitations from the equilibrium point for three source resistances $R=1\,\Omega$,$R=1.2\,\Omega$, and $R=1.8\,\Omega$ with $L=0.1\,\upmu$H and $C=1\,\upmu$F.
The eigenvalues are all complex-valued and negative for $R=1\,\Omega$ and $R=1.2\,\Omega$, but positive for $R=1.8\,\Omega$. The nonlinear ODE in Eq.~\ref{eqn:sdflsödkfjg} was solved numerically. For the unstable condition, the arc extinguishes and approaches the dark current point. Considering the simple stability analysis, it is a reasonable assumption that the power supply and the reactive elements form a coupled dynamic system. As the negative slope of the arc cannot be freely modified, Eq.~\ref{eqn:skljshdf} implies that the source resistance should overcompensate the negative differential resistance of the arc, but at the same time $L/(RC)$ should also be large. Consequently, the capacitance should be low, while the inductance should be high. In the underdamped case, the square root in the eigenvalues in Eq.~\ref{eqn:sdlskdjf2} is purely complex such that the time characteristic time constant is given in physical time by 
\begin{align}
    \tau_0=\frac{L}{R}\frac{2}{\frac{L}{R^2C}+\frac{1}{R}\frac{dU}{dI}}.
\end{align}
The time constant is dominated by $L/R$, but the closer the condition in Eq.~\ref{eqn:skljshdf} is met, the longer it will be. Increasing inductance has a smoothing effect on the electrical variables, which will in turn also influence the arc stability. Therefore, a buck converter with a comparably large output inductor, as will be discussed in more detail in Sec.~\ref{sec:PSU}, is ideally suited to suppress unwanted dynamics.

\section{Measurement Setup}
\subsection{Measurement Cell}
\begin{figure}[htbp]
	\centering
	\includegraphics[width=0.45\textwidth]{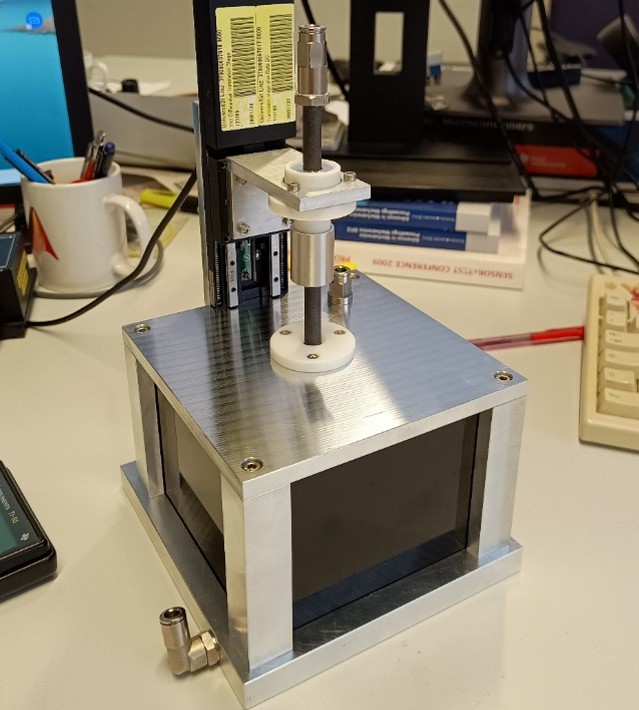}
    \includegraphics[width=0.464\textwidth]{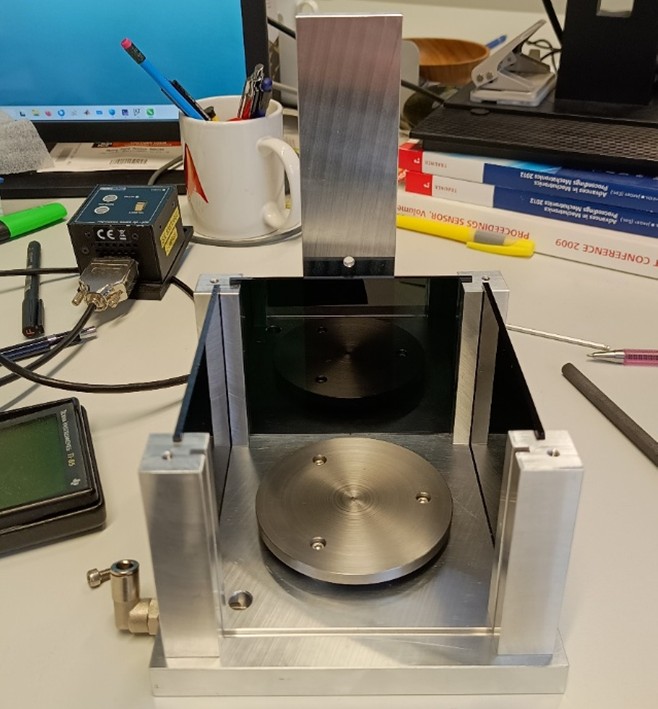}
	\caption{First demonstrator cell. The arc can be viewed through protective welding glasses from all sides.} \label{fig:fig6}
\end{figure}

The measuring chamber in Fig.~\ref{fig:fig6} was designed to precisely investigate the properties of the arc. The four sides of the chamber are equipped with DIN6 welding glasses of standard dimensions $90\,$mm$\times 110$\,mm. These are inserted tightly into the grooves of the aluminium parts by means of Teflon tape and offer protection against glare and UV radiation. A linear stage moves the electrode in the vertical direction through a heat-stable Teflon bushing in the top plate. The electrode has a negative potential against the aluminium housing and is additionally electrically insulated by the Teflon element. It is designed accordingly so that there is no flashover from the electrode to the cover plate when the arc is stretched. The round aluminium element below the electrode holder in Fig.~\ref{fig:fig6} (left) is a spring element. This limits the contact pressure of the electrode during contact ignition. On the one hand, this serves as mechanical protection of the electrode, but it is also used to create a comparatively high contact resistance by means of light contact pressure, which allows the contact point to be preheated before plasma ignition. The graphite electrode has a nominal diameter of 8\,mm and a length of 200\,mm. The bore through which the gas can be supplied has a diameter of 2\,mm. In the following tests, however, solid electrodes were used. In these tests, the gas is supplied via the gas connection on the lid plate (rear left in Fig.~\ref{fig:fig6} (left)) or via the base plate (bottom right). Fig.~\ref{fig:fig6} (right) shows the open measuring chamber. The round contact plate is made of structural steel St52 and is thermally decoupled from the base plate by means of ceramic spacers, but is electrically connected to the housing via three M3 screws. Initial tests were carried out using the laboratory power supply unit HP6269B (thyristor stage with downstream linear regulator, 0-40V, 0-50A). 

\begin{figure}[htbp]
	\centering
	\includegraphics[width=1\textwidth]{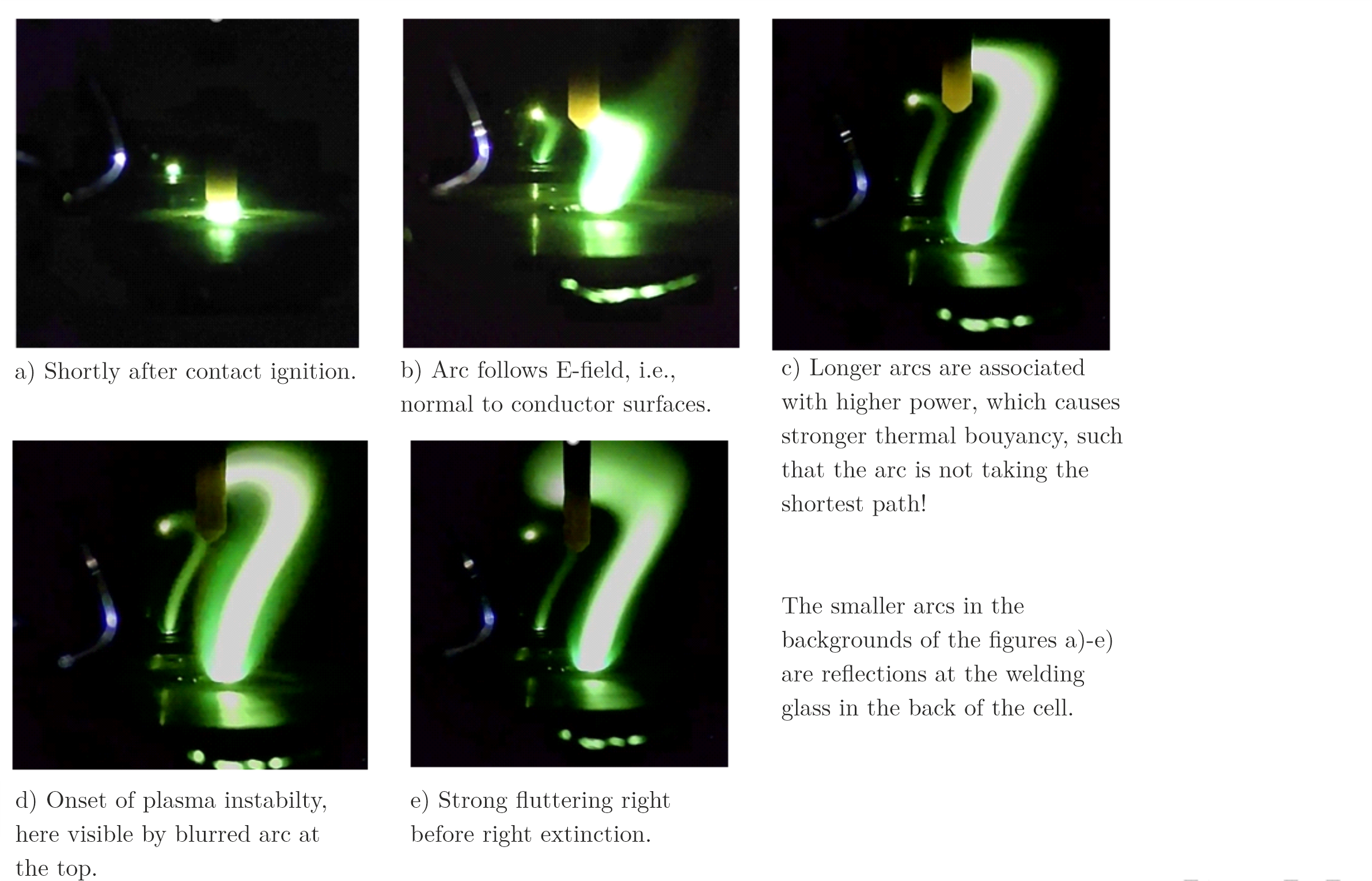}
	\caption{Electric arcs at varying electrode gaps using a conventional laboratory power supply (40V/50A) in constant-current mode at 20A. Longer arcs are accompanied by higher thermal energies, which promote instability. The plasma fluctuations shortly before extinction are associated with strong oscillations of the supply circuit.} \label{fig:fig7}
\end{figure}
 In the first experiment, it was observed that contact ignition was not possible with the directly connected PSU. Adding a laboratory inductor of 100mH inductance solved the problem. At that time, it was suspected that the inductor would force the current flow by generating a voltage spike shortly after leaving contact. Surprisingly, no voltage spike was observed, and it was sufficient to replace the inductor with an ohmic resistor with the same value as the series resistance of the inductor of approx. 1$\Omega$. In retrospect, it is clear that the stability condition in Eq.~\ref{eqn:sdfsdjfhgsd} can only be satisfied by adding the series resistor. Figure~\ref{fig:fig7}(a) to (e) shows the evolution of the arc beginning with contact ignition (a) followed by stretching. For the experiment, the PSU was used in constant current mode, set to 20A.  It can be observed in (b) that the arc contacts the electrode and the steel plate in normal direction, i.e., following the electric field. The arc is additionally stretched by the thermal buoyancy, which even causes the arc to climb along the shaft (c). A longer arc at constant current implies higher voltage and consequently higher arc power. This causes forces due to turbulent gas flows to dominate over the roughly constant electrical forces. In the higher power ranges, strong oscillations of the arc accompanied by strong oscillations of power supply could be observed (see Fig.~\ref{fig:fig7}(e)).  
\subsection{Power Supply}\label{sec:PSU}

Due to the above-mentioned limitations, a current regulator based on a buck converter was implemented, which is downstream of the laboratory power supply. This topology (see Fig.~\ref{fig:fig8}) is particularly suitable for this task, as the output inductance already inherently stabilizes the output current for a short time. The input variable of the controller is a pulse-pause modulated (PWM) signal with duty cycle denoted by  $p=[0\dots 1]$. The output voltage $U_A$ for input voltage $U_B$ and duty cycle $p$, neglecting the diode and the emitter-collector drop of the IGBT, is given by   
\begin{equation}
    U_A=p\times U_B\label{eqn:sdfgskldfh1}.
\end{equation}
This means that a superimposed current regulator must adjust $p$ according to the dynamic load. 
The ripple current $\Delta I$ (i.e., the current change due to charging and discharging of the inductor during one switch cycle) and time constant $\tau$ of the output current are given by
\begin{align}
    \Delta I_A = \frac{U_B(1-p)p}{L f_\text{osc}}\label{eqn:sfksjdfg1},\\
    \tau=L/\left.\left|\frac{dU}{dI}\right|\right|_\text{o.p.}\label{eqn:sfksjdfg2}.
\end{align}
A high switching frequency is required to achieve low current ripple while maintaining sufficiently fast current-control dynamics. The IGBT of type IXXN100N65CH4H1 from IXYS is suitable for switching frequencies up to 60 kHz, so a switching frequency = 50 kHz is used. The inductance is of the type Hammond 195C75 with $L=2.5\,$mH and $R_s$=8.6\,m$\Omega$. The fast switching is required because the buck converter is electrically a voltage source (see Eq.~\ref{eqn:sdfgskldfh1}) with low output resistance, which would not overcompensate the negative resistance of the electric arc. Adding an appropriate series resistor is not a viable option when energy efficiency has to be taken into consideration. The current-controlled buck converter does only appear as a current source to the load when the controller reacts much faster than the time constant in Eq.~\ref{eqn:sfksjdfg2}, which is roughly 2ms. As a rule of thumb, the PWM frequency should be 5 times higher than the update rate of the controller. 
\begin{figure}[htbp]
	\centering
	\includegraphics[width=1\textwidth]
	{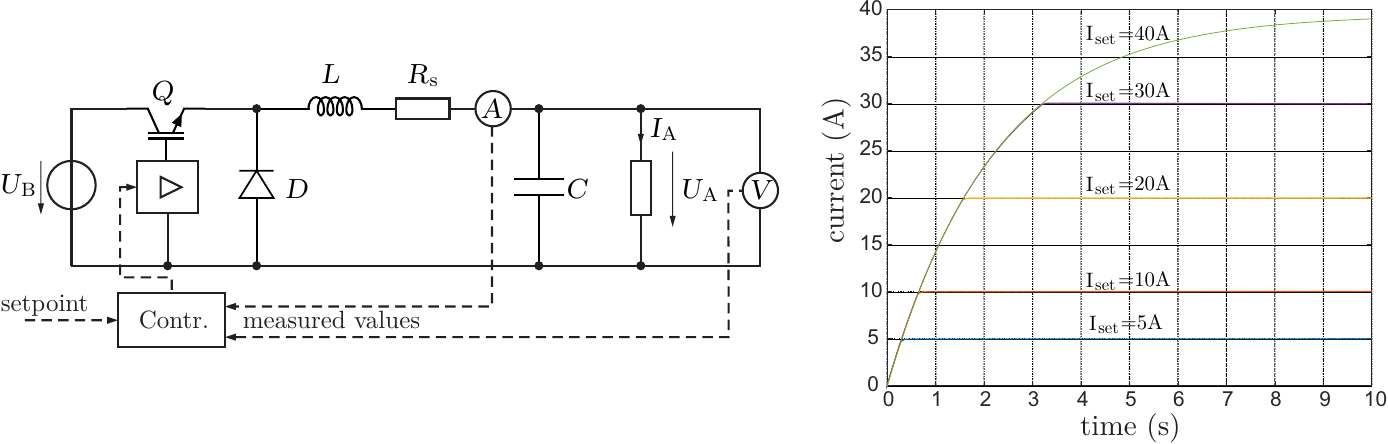}
	\caption{Left) Circuit diagram of the buck converter. Right) SPICE-simulated current-setting characteristic for various current setpoints at a 1$\Omega$ load. The time constant is $L/R$} \label{fig:fig9}
\end{figure}
Figure \ref{fig:fig9} (right) shows SPICE simulations of the control behavior at a constant load of $R=1\Omega$. A 
simple PI controller was used for this test. The target current jumps from 0A to 5A, 10A, 20A, 30A, and 40A follow with the constant time given in Eq.~\ref{eqn:sfksjdfg2}. The ripple currents =40mA, 65mA, 86mA, and 62mA determined in the simulation agree with the values given by Eq.~\ref{eqn:sfksjdfg1}. The maximum value occurs at $p=0.5$ (20A). 

\begin{figure}[htbp]
	\centering
	\includegraphics[width=1\textwidth]{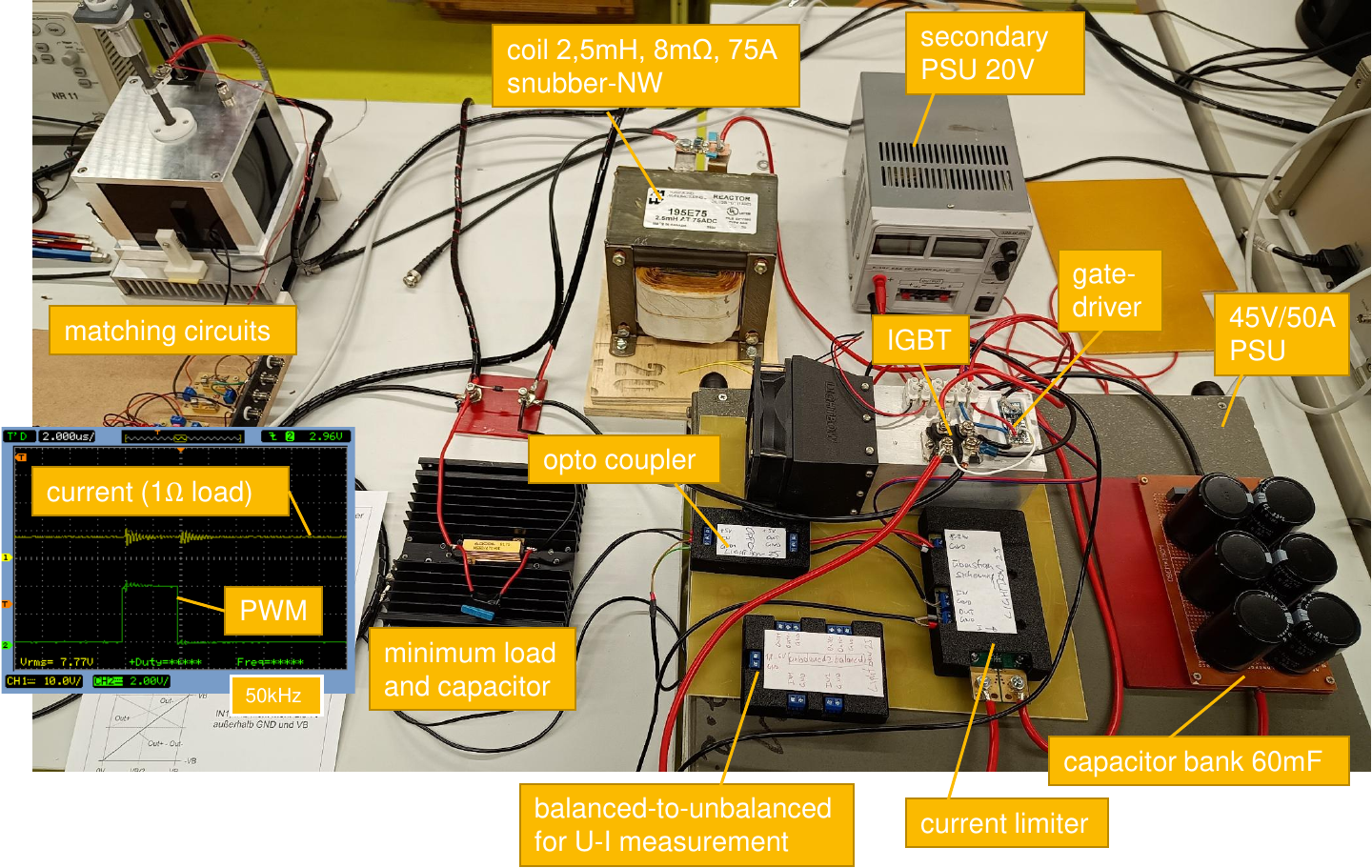}
	\caption{Measurement setup with electronic components. The control system (not shown) measures the current using a fluxgate sensor and adjusts the PWM of the buck converter according to a setpoint.} \label{fig:fig8}
\end{figure}
To control the actual electric arc current, the PI controller is replaced by an adaptive state controller with disturbance compensation described briefly in Sec.~\ref{REG} and in detail in~\cite{Lukas}. \\

Figure~\ref{fig:fig8} shows the electrical components of the experimental setup. The capacitor bank (60mF) is used to prevent voltage spikes from triggering the power supply's surge shutdown when the IGBT is switched off. In order to protect the IGBT from high current peaks that can be caused by the capacitors in short-circuit conditions, an overcurrent shutdown circuit using a Hall current sensor with a fast response time ($<2\upmu$s) and hysteresis was implemented. 

In the case of contact ignition, the graphite electrode heats up, which facilitates the release of the electrons required to generate the plasma. The potential of the graphite electrode is therefore negative compared to the base plate. The regulator in this form delivers a positive voltage, which is why galvanic isolation devices are also used for the control signal and the measured quantities.

\section{Control system and implementation}\label{REG}
An industrial process computer is used to implement the current control, on which the control algorithms are executed in the real-time environment TwinCAT from Beckhoff. The controller operates with a fixed sampling time of 125$\upmu$s, which achieves a sufficiently high temporal resolution to precisely record and compensate for the fast dynamic effects of the arc and the power electronics. In particular, the short cycle time enables the targeted damping of those oscillations that result from the interaction of the negative differential conductivity of the arc and the parasitic energy storage systems in the supply path. The current controller is designed as an adaptive state-feedback controller with disturbance compensation. The aim of the control is to precisely specify the arc current despite strongly varying load characteristics. The arc represents a nonlinear and time-variant load, whose effective differential resistance changes with both the arc length and the local plasma temperature. In addition, the arc voltage acts as the dominant disturbance variable, as it is directly incorporated into the dynamics of the inductor current change, but can be compensated for by appropriate pilot control. The compensation prevents changes in the arc length from leading directly to current deviations. The disturbance compensation therefore improves both reference-tracking performance and robustness against abrupt changes in the plasma state, for example during the transition between metastable arc lengths or in the case of short-term arc displacements on the bath surface/metal plate. The controller adapts its parameters to changes in differential arc resistance, which allows constant dynamics to be achieved over a wide working range of typically 5\,A to 40\,A. The resulting closed current control section exhibits a well-damped behavior with a short settling time and allows stable adjustment even under strongly fluctuating plasma conditions. The hardware-related connection of the power electronics is carried out via a microcontroller platform based on the Texas Instruments F28P65 (C2000 series). The microcontroller is operated as an EtherCAT slave and provides the time-critical input and output units. These include, in particular:
\begin{itemize}
\item Generation of the PWM signal for the IGBT of the step-down controller with high temporal resolution 
\item Detection of arc voltage by means of fast analog-to-digital conversion 
\item Measurement of inductor current via a fluxgate-based current sensor 
\item Deterministic communication with the higher-level process computer via EtherCAT 
\end{itemize}

By outsourcing the time-critical signal processing to the microcontroller, very low dead times can be achieved. At the same time, the EtherCAT communication enables synchronous integration into the real-time system of the process computer. The resulting overall structure combines the high computing power and flexibility of the TwinCAT environment with the fast peripheral connection of the C2000 microcontroller platform. The selected architecture also allows for easy expansion with additional control functions, such as superimposed power control or model-based estimation of the arc length based on the electrical signals. Overall, it can be seen that the combination of fast real-time control, interference compensation, and high-performance power electronic actuators can achieve a significant improvement in current stability.

\section{Instrumentation}
\begin{figure}[htbp]
	\centering
\includegraphics[width=1\textwidth]{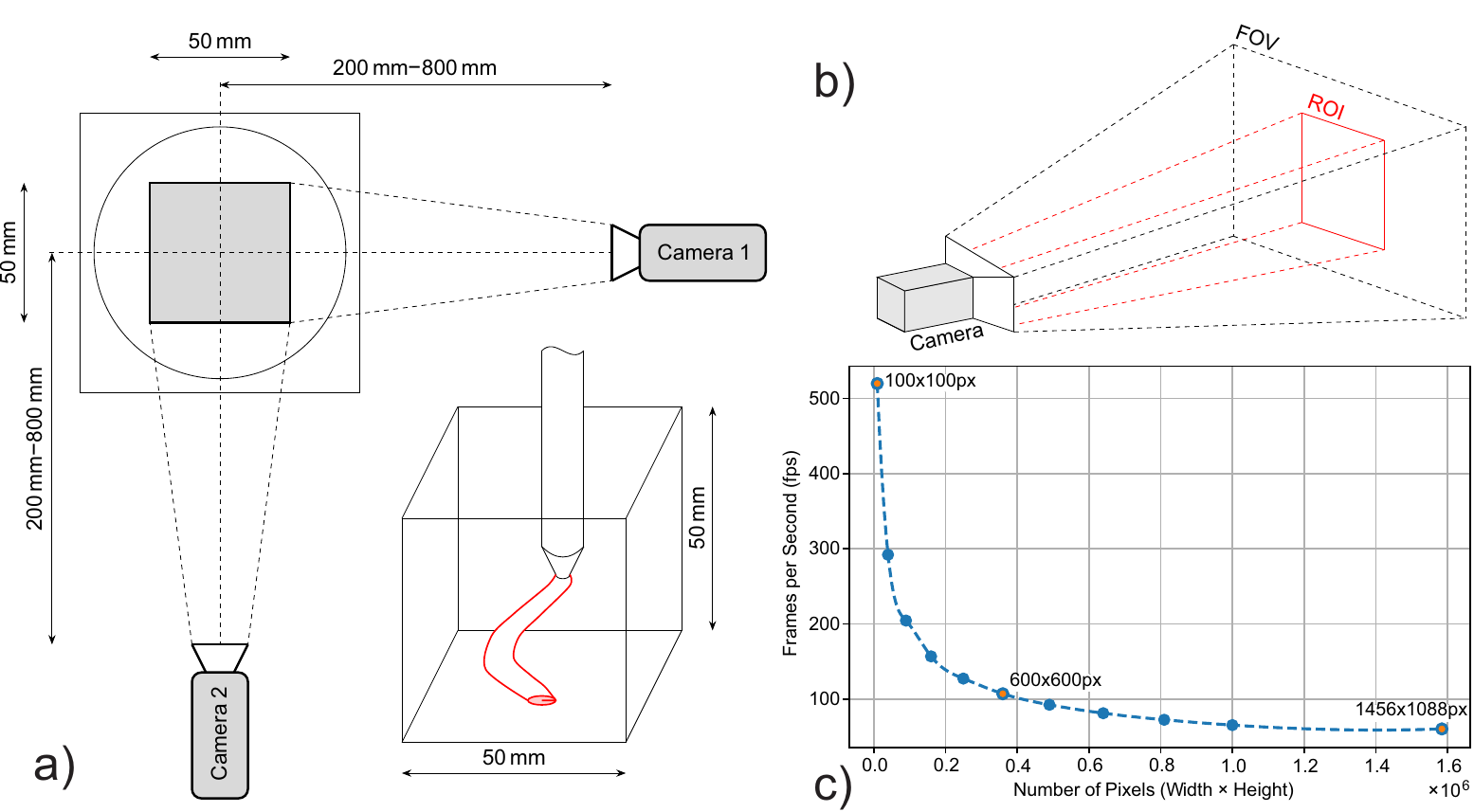}
    \caption{a) Physical requirements matched to a concept plan for the measurement setup. b) Illustration of the difference between the field of view (FOV) and the region of interest (ROI). c) Visualization of the dependency of the framerate (frames per second) over the read sensor pixels (ROI).
} \label{fig:fig10}
\end{figure}

\begin{figure}[htbp]
	\centering
\includegraphics[width=1\textwidth]{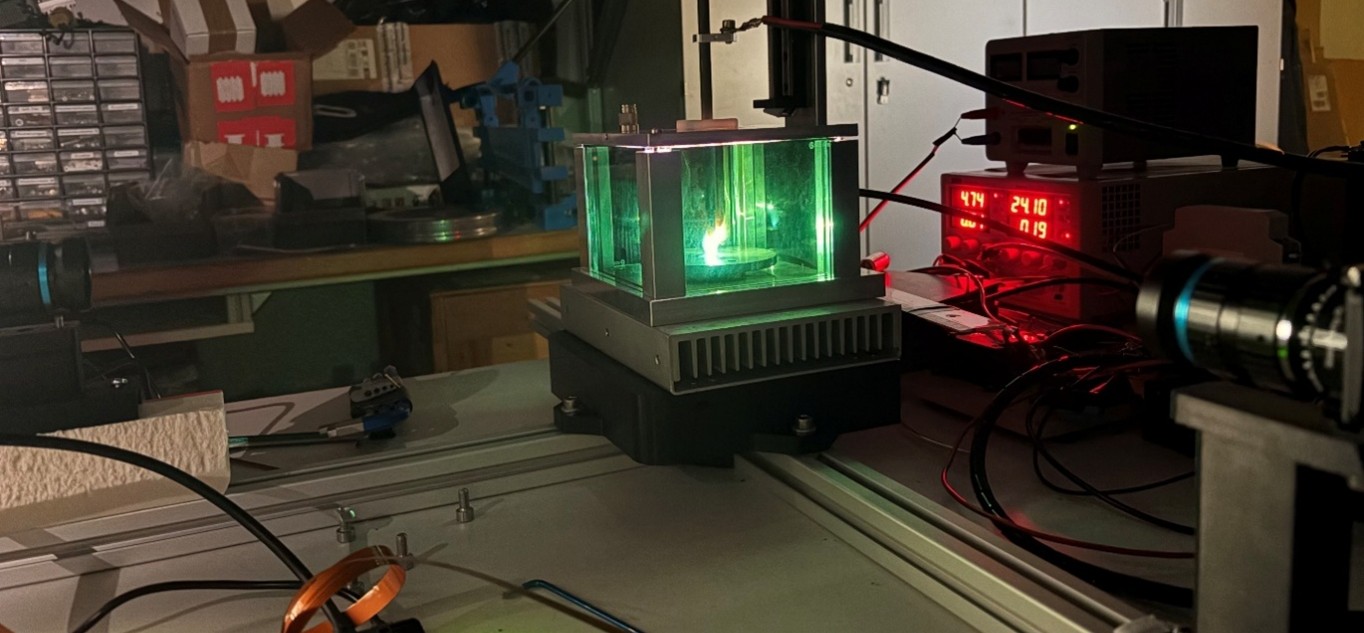}
\includegraphics[width=1\textwidth]{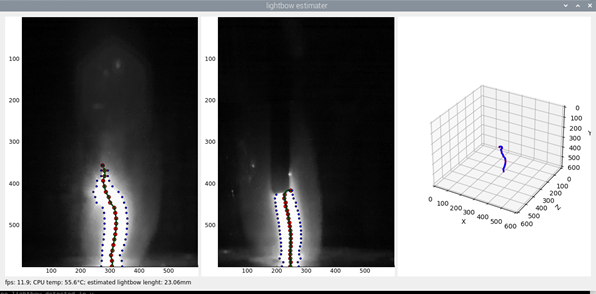}
    \caption{ Stereo camera system setup (above) and live view of the arc reconstruction (below). The update rate of the estimated lengths and energies increases to approx. 110Hz when the visualization is deactivated } \label{fig:fig11}
\end{figure}

\begin{figure}[htbp]
	\centering
	\includegraphics[width=0.68\textwidth]{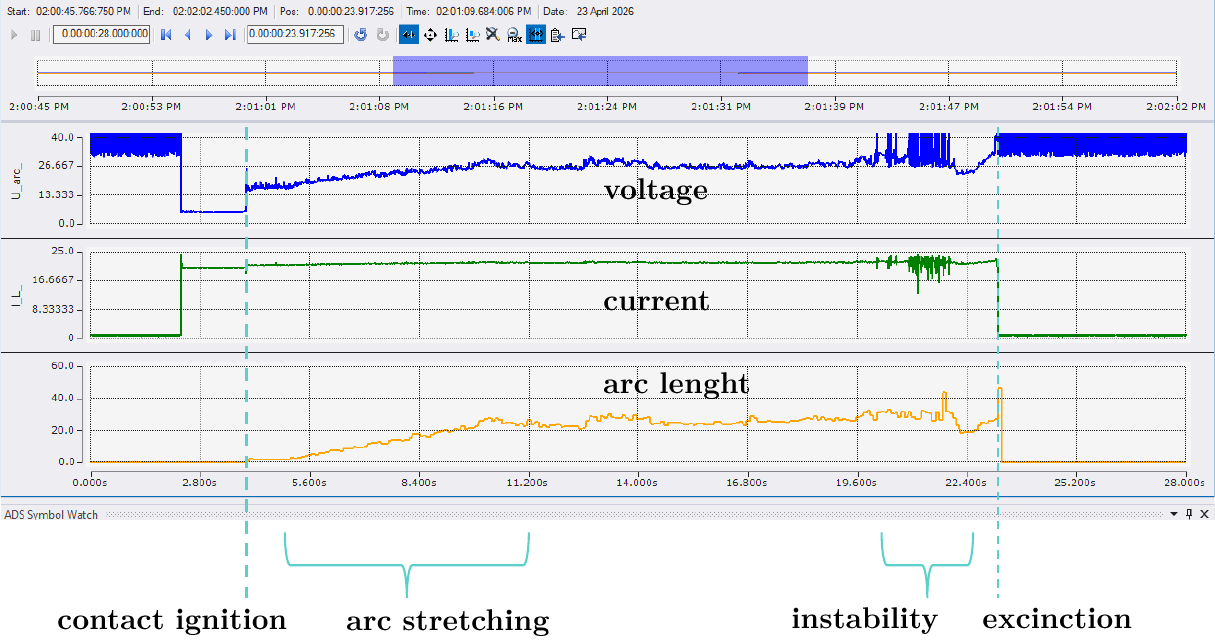}
	\caption{Arc ignition, stretching, and extinction for a supply voltage of 40V with current controller enabled. The measured arc length follows the voltage as expected. The arc extinguishes when the current controller requires more voltage than is provided by the supply voltage.} \label{fig:fig12}
\end{figure}

To characterize arc stability, a stereoscopic system in Fig.~\ref{fig:fig10}a was implemented~\cite{Lehner}. 
Arc stability is defined as the temporal stability of the arc length. In the target application, the optical observability of the arc is usually not given, which is why the test bench is used to investigate the extent to which the electrical signals correlate with the arc length. The cameras used are 1.6-megapixel global shutter with a Sony IMX296 sensor for Raspberry Pi, with a 16 mm telephoto lens. The global shutter causes a simultaneous exposure of all pixels of the sensor. The camera type allows a region-of-interest (ROI, see Fig.~\ref{fig:fig10}b) to be defined, in which only one image area has to be read, which significantly increases the frame rate. The frame rate versus resolution is shown in Fig.~\ref{fig:fig10}c. The target frame rate of 100 Hz is achieved using a \(600\times600\) pixel ROI. The object distance and the aperture setting were chosen in such a way that a cube-shaped area with a side length of 50\,mm can be sharply imaged. The stereo images are recorded synchronously on a Raspberry Pi 5. The arc is identified in two orthogonal planes and interpolated using spline functions (see Fig.~\ref{fig:fig11} lower figure). The length is calculated from the centerline of the arc. The measurement method was calibrated with bent wires of known length. The measurement data are transmitted to the process computer at an update rate of > 100Hz. Fig.~\ref{fig:fig12}, shows the synchronous live data visualization in TwinCAT from Beckhoff.  

\section{Results}

 \begin{figure}[h]
 	\centering
 	\includegraphics[width=0.8\textwidth]{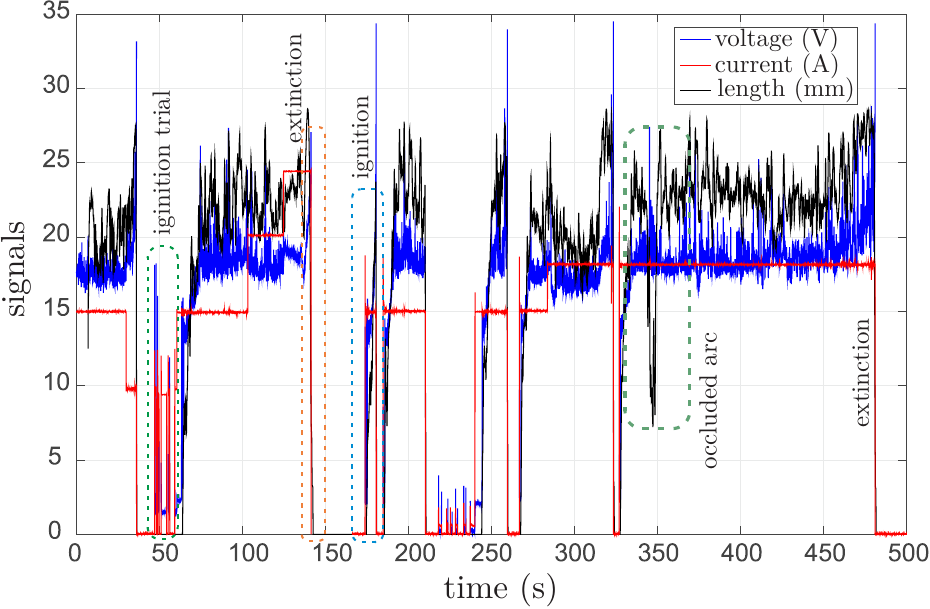}
 	\caption{Current, voltage and estimated length of the arc. } \label{fig:fig13}
 \end{figure}

\begin{figure}[htbp]
	\centering
	\includegraphics[width=0.86\textwidth]{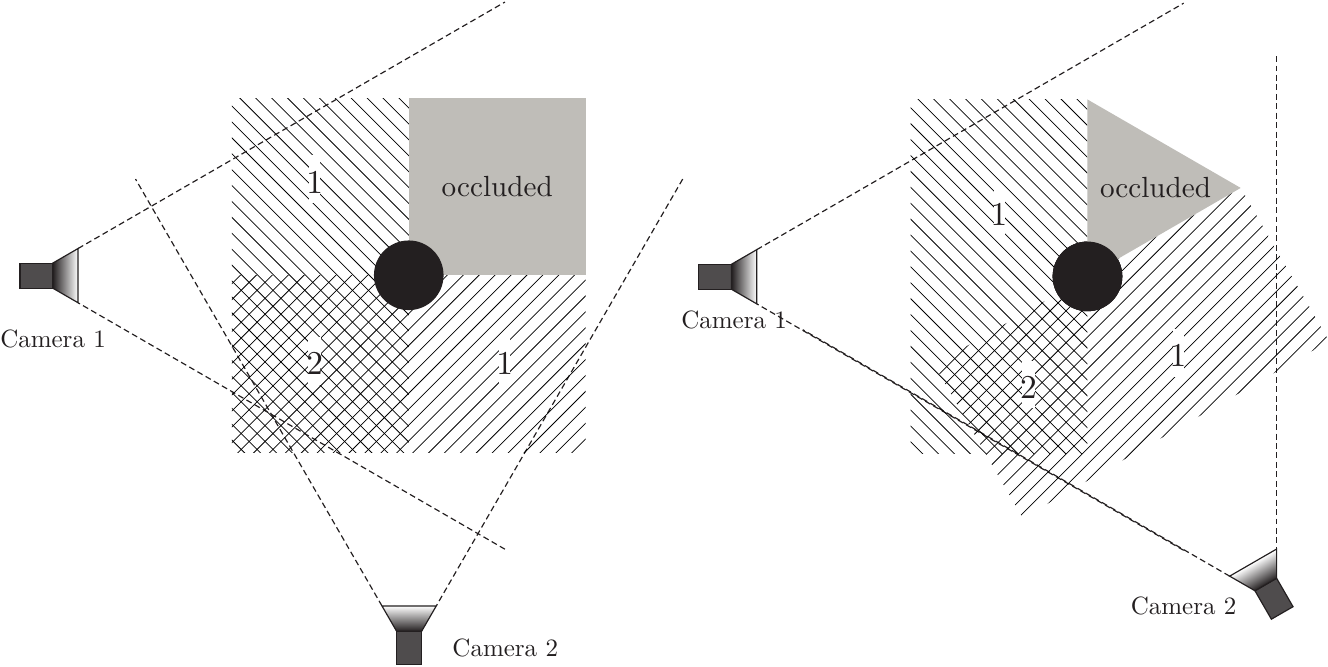}
	\caption{Left) A climbing arc is occluded by the electrode in both camera images with a chance of 25\% and with a chance of 75\% in at least one image. Right) Using a skewed camera configuration (here 120$^\circ$) reduces the chance of occlusion in both images but also reduces the chance that it is visible in both images.} \label{fig:fig14}
\end{figure}

Figure~\ref{fig:fig13} shows measured currents, voltages, and arc lengths. After contact ignition, the arc is stretched by increasing the gap between electrode and the ground plate. The current regulator was set to values in the range of 10-22\,A. Current setpoint changes occurred consistently with a rise time (10-90\%) of approx. 4ms. The stability of the current controller was assessed within the interval from 285s to 318s. In this section, the current was constant 18A with a coefficient of variation (standard deviation/mean) of 0.03\%. The length drop around 350s is an artifact, as the arc climbed along the shaft of the electrode (similar to Fig.~\ref{fig:fig7}c) and was occluded by the electrode in both camera images. In the case of arc climbing, the chance that the arc is occluded in at least one image is approx. 75\%. To reduce this problem, the algorithm detects if the heights of the arcs are the same in both images. If not, the algorithm extrapolates the arc in the image with the shorter height to match the height of the longer arc, under the constraint to enter the shaft in the normal direction. This reduced the chance of full occlusion to approx. 25\%, as shown in Fig.~\ref{fig:fig14}.  
Changing the angle of the cameras, shown in Fig.\ref{fig:fig14} (right), reduces the chance of full occlusion, but also the chance of visibility in both images, such that the length extrapolation must be more reliable in such a configuration. Due to the cubic shape of the measurement cell in Fig.~\ref{fig:fig6}, the angles were not adapted in our experiments. 

\begin{figure}[htbp]
	\centering
	\includegraphics[width=0.49\textwidth]	{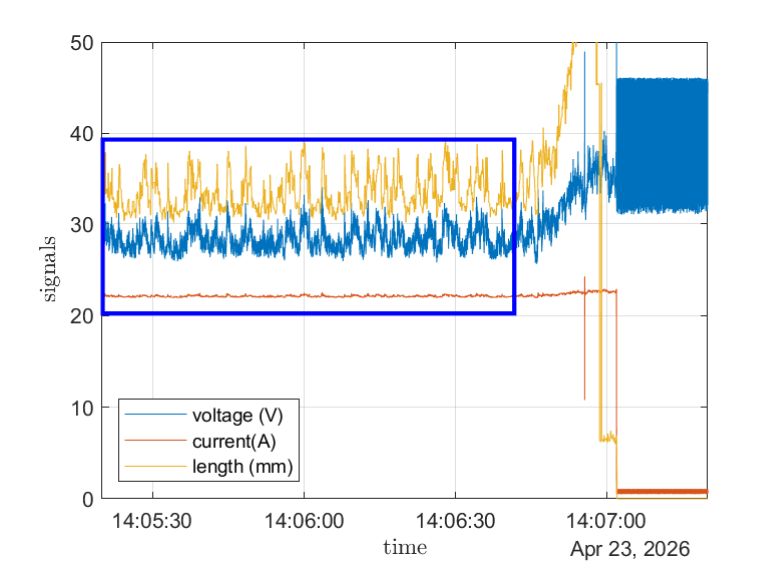}
    \includegraphics[width=0.49\textwidth]	{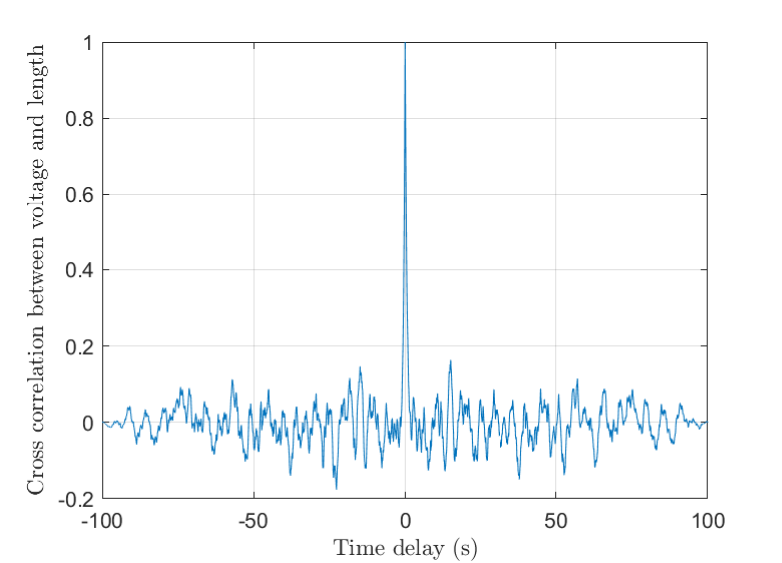}
	\caption{Left) Segment of a relatively stable arc condition marked by the box. Right) A strong cross-correlation between arc voltage and measured length could be identified. Therefore, the variance of voltage under constant current condition is a clear indicator of arc stability.} \label{fig:fig15}
\end{figure}

However, using the proposed or a similar camera system in the industrial process will not be possible due to various limitations, including that the arc cannot be viewed from the sides, and dust will obstruct the view. Therefore, it is of interest whether the arc stability can be determined by the electrical signals as suggested by Fig.~\ref{fig:fig1} (right), where arc elongation is associated with higher voltages. To answer this question, the normalized cross-correlation function of arc voltage and arc length at stable conditions shown in Fig.~\ref{fig:fig15} (left) is calculated. It is found that the correlation between both quantities is strong, such that the voltage variance at constant current is a clear indicator of arc stability. 

\section{Conclusions and Outlook}
The electrical arc is a highly stochastic system. The power supply has an impact on the arc stability due to the negative differential resistance of the arc, particularly when reactive elements such as wire inductances and capacitances are considered. It was shown that a fast current-controlled buck converter is ideally suited to provide stable arc current due to its comparably large output inductance, which has a stabilizing effect. A fast stereoscopic camera system was designed to assess the arc stability by measuring the arc length variation. It could be shown that under constant current conditions provided by the current controller, the arc voltage is closely correlated with the arc length variation. Therefore, in the industrial target process, arc stability may solely be assessed by analyzing the electrical signals, and dynamic supervisory current controller can contribute to arc stability. 

\begin{figure}[htbp]
	\centering
	\includegraphics[width=0.4\textwidth]	{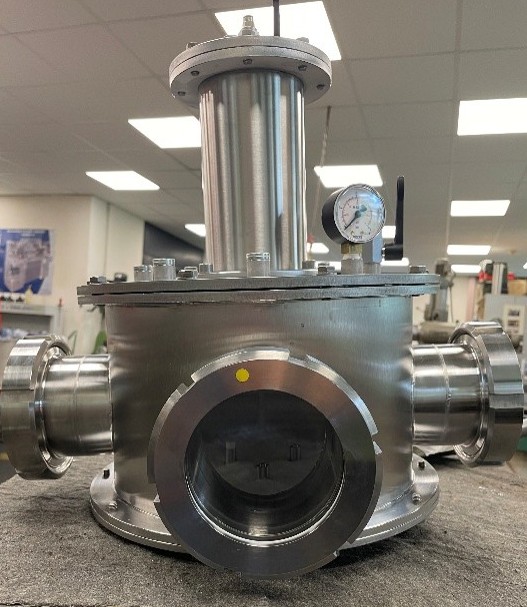}
    \includegraphics[width=0.5\textwidth]	{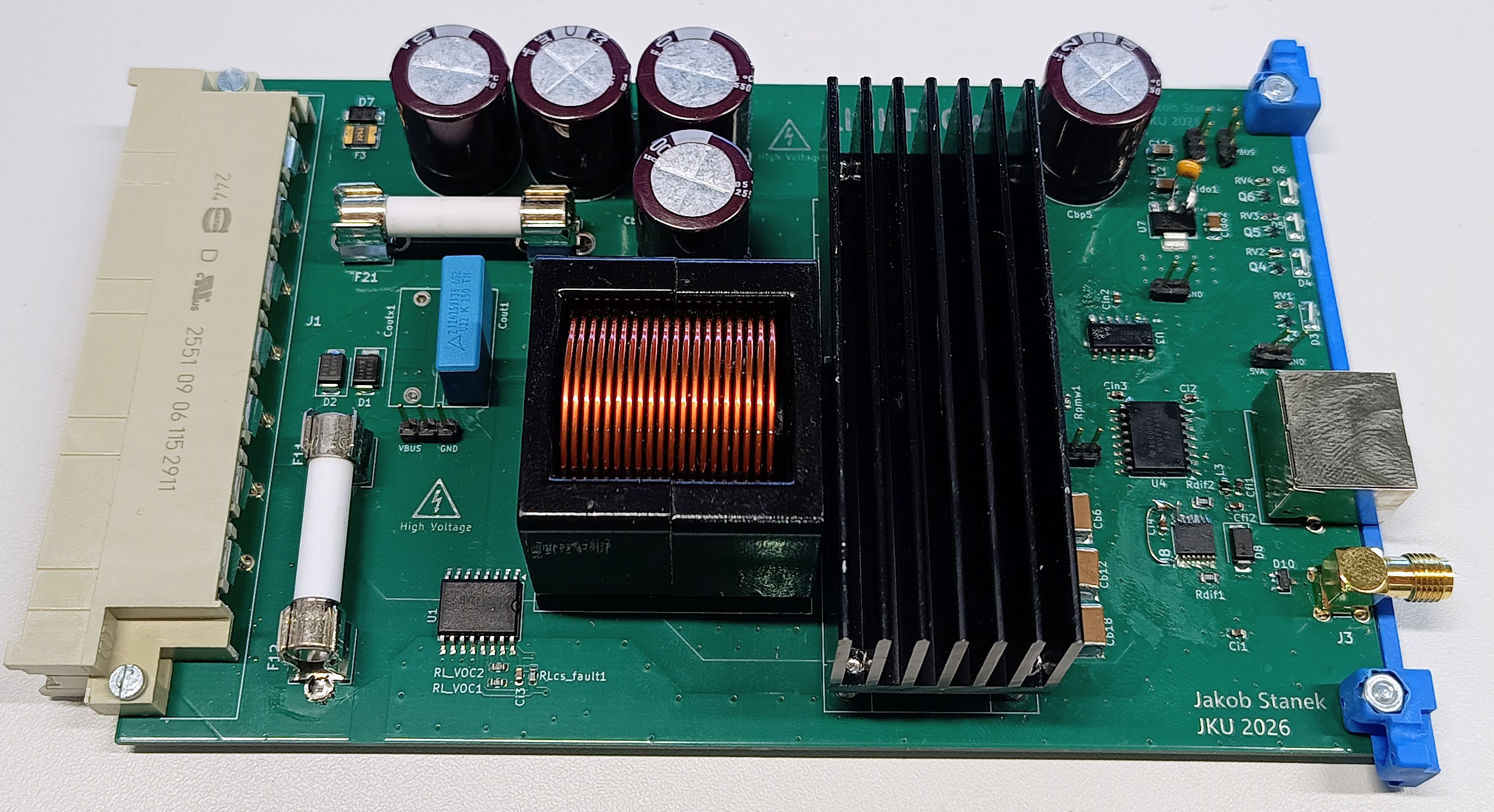}
	\caption{Left) Redesigned reactor demonstrator. Right) Prototype of GaN FET-based buck converter modules designed for multiphase operation in 19-inch rack chassis~\cite{Stanek}.} \label{fig:fig16}
\end{figure}
A larger measuring chamber, shown in Fig.~\ref{fig:fig16} (left), with a nominal thermal power of 5KW is currently being implemented. This already has comparable dimensions to a metallurgical test plant, so that the measurement and control strategies researched here can be transferred. The power section is being replaced by a specially developed, arbitrarily parallelizable, GaN-FET-based power controller~\cite{Stanek}. The individual modules are operated in multiphase (parallel but with phase-shifted PWM signals) with a switching frequency of up to 500\,kHz. This promises even better control over the electrical characteristics of the electro-thermal system "arc".

\section*{Acknowledgement}
The LIGHTBOW project was funded by the Austrian Research Promotion Agency (FFG) as part of Expedition Zukunft Wissenschaft. E.K.R. acknowledges additional funding from the FFG COMET Module PlasmArc4Green to the COMET Center K1-MET.


\begin{thebibliography}{00}
\bibitem{Hiebler}
Hiebler, H., and  Plaul, J. F. (2004). Hydrogen plasma smelting reduction-an option for steelmaking in the future. Metalurgija, 43(3), 155-162.
\bibitem{Pauna}
Pauna, H., Ernst, D., Zarl, M., Aula, M., Schenk, J., Huttula, M., and Fabritius, T. (2022). Hydrogen plasma smelting reduction process monitoring with optical emission spectroscopy–Establishing the basis for the method. Journal of Cleaner Production, 372, 133755.
\bibitem{Adami}
Adami, B., Farkas, M. A., Schenk, J., and Zarl, M. A. (2026). Advancing hydrogen plasma smelting reduction: Experimental insights from a pilot plant. Matériaux and 
Techniques, 114(1), 105.
\bibitem{Zarl1} Zarl, M. A., Farkas, M. A., and Schenk, J. (2020). A study on the stability fields of arc plasma in the HPSR process. Metals, 10(10), 1394.
\bibitem{Leitner}
Leitner, E. M. (2023). Methodenentwicklung zur Stabilisation des Elektrischen Lichtbogens im Wasserstoff-Argon Schmelzreduktionsplasma. M.Sc. thesis, Montanuniversität Leoben 
\bibitem{Zarl2}
Ernst, D., Zarl, M. A., Farkas, M. A., and Schenk, J. (2023). Effects of the Electrodes’ Shape and Graphite Quality on the Arc Stability During Hydrogen Plasma Smelting Reduction of Iron Ores. steel research international, 94(7), 2200818.
\bibitem{Reichel}
Reichel, E., Ecker, L., Voglhuber-Brunnmaier, T., Schneiderbauer, S., Quick, C., Jelinek, A., and Zarl, M. A. (2026). Stability Enhancement of DC Arcs in Reducing Argon-Hydrogen Atmospheres via Current Regulation. In \textit{EEC 2026 -- 14th European Electric Steelmaking Conference and EMECR 2026 -- 5th International Conference on Energy and Material Efficiency and CO2 Reduction in the Steel Industry}, Milan, Italy, May 11--13, 2026.
\bibitem{Rieder}
Rieder, W. (1967). Plasma und Lichtbogen, Friedr. Vieweg und Sohn, Braunschweig. 
\bibitem{Andrea}
Andrea, J., Besdel, P., Zirn, O., and Bournat, M. (2015, November). The electric arc as a circuit component. In IECON 2015-41st Annual Conference of the IEEE Industrial Electronics Society (pp. 003027-003034). IEEE.
\bibitem{Lukas}
Ecker, L., and Voglhuber-Brunnmaier, T. (2026). Adaptive Average Current-Mode Control of a Buck Converter Supplying Plasma Arc Loads. ArXiv. https://arxiv.org/abs/2610.07930
\bibitem{Strogatz}
Strogatz, S. H. (2024). Nonlinear dynamics and chaos: with applications to physics, biology, chemistry, and engineering. CRC press.
\bibitem{Lehner}
Lehner, B. (2026). Stereoscopic Characterization of Electric Arcs: Real-Time Visualization and Feature Extraction, M.Sc. thesis, Institute for Microelectronics and Microsensors, Johannes Kepler University, Linz
\bibitem{Stanek}
Stanek, J. (2026). Modularer Buckconverter zur Lichtbogensteuerung, B.Sc. thesis, Institute for Microelectronics and Microsensors, Johannes Kepler University, Linz
\end{thebibliography}
\end{document}